\documentclass[reqno, natbib]{amsart}

\allowdisplaybreaks
\usepackage{natbib,latexsym,amsmath,amssymb,color,enumerate,graphicx,url}

\numberwithin{equation}{section}

\usepackage[]{hyperref}
            
\hypersetup{colorlinks,
            linkcolor=blue,
            anchorcolor=blue,
            citecolor=blue}

\begin{document}

\title[Network Sessions]{Extremal Dependence Analysis of Network Sessions}

\author[L. L\'opez-Oliveros]{Luis L\'opez-Oliveros}
\address{Luis L\'opez-Oliveros\\
Department of Statistical Science\\
Cornell University \\
Ithaca, NY 14853}
\email{ll278@cornell.edu}

\author[S.I.\ Resnick]{Sidney I.\ Resnick}
\address{Prof. Sidney Resnick\\
School of Operations Research and Information Engineering\\
Cornell University \\
Ithaca, NY 14853}
\email{sir1@cornell.edu}

\thanks{Luis L\'opez-Oliveros'  research was partially supported by CONACyT (Mexican Research Council of Science and Technology) Contract 161069.
Sidney Resnick's research was partially supported by
ARO Contract W911NF-07-1-0078 at Cornell University.
}

\begin{abstract}
We refine a stimulating study by \cite{sarvotham:riedi:baraniuk:2005}
which highlighted the influence of peak transmission rate on network
burstiness. From TCP packet headers, we amalgamate packets into
sessions where each session is characterized by a 5-tuple
$(S,D,R,R^\vee,\Gamma)$=(total payload, duration, average transmission rate, peak
transmission rate, initiation time). After careful consideration, a new definition of
peak rate is required. Unlike \cite{sarvotham:riedi:baraniuk:2005} who
  segmented sessions into two groups labelled alpha and beta, we
  segment into 10 sessions according to the empirical quantiles of the
  peak rate variable as a demonstration that the beta group is far
  from homogeneous. Our more refined segmentation reveals
  additional structure that is missed by segmentation into two groups.
In each segment, we study the dependence structure of $(S,D,R)$ and find that it varies across the groups. Furthermore, within each segment,
session initiation times are well approximated by a Poisson process
whereas this property does not hold for the data set taken as a whole.
Therefore, we conclude that the
peak rate level is important for understanding structure and
for constructing accurate simulations of data in the wild. 
We outline a simple
method of simulating network traffic based on our findings.

\end{abstract}

\maketitle

\section{Introduction}\label{sec:intro}

Statistics on data networks show empirical features that are
surprising by the standards of classical queuing theory. Two
distinctive properties, which are called 
 \textit{invariants} in the network literature, are:
\begin{itemize}
\item Heavy tails for quantities such as file sizes
  \citep{arlitt:williamson:1996,willinger:paxson:1998,leland:taqqu:willinger:wilson:1994,
  willinger:paxson:taqqu:1998a},
  transmission durations and transmission delays 
 \citep{maulik:resnick:rootzen:2002, resnick:2003}.
\item Network traffic is bursty \citep{sarvotham:riedi:baraniuk:2005}, with rare but influential periods of high transmission rate punctuating typical periods of modest activity. Burstiness is a somewhat vague concept but it is very important in order to understand network congestion.
\end{itemize}

When studying burstiness, bursts are observed in the sequence of
bytes-per-time or packets-per-time, which means that a window
resolution is selected and the number of bytes or packets is counted
over consecutive windows. \cite{sarvotham:riedi:baraniuk:2005} attempt
to explain the causes of burstiness at the user-level. By the
user-level, we mean the clusters of bytes that have the same source
and destination network addresses, which we term \textit{sessions}. As
a simplification, associate a session with a user downloading a file,
streaming media, or accessing websites; a more precise definition is
given later. For each session, measurements are taken on the size or
number of bytes transmitted, the duration of the transmission and the
average transfer rate. If the primary objective is to explain
sources of burstiness, the session peak rate arises as a natural
additional variable of interest. The peak rate is computed as the
maximum transfer rate over consecutive time slots.

In order to explain the causes of burstiness at the user-level,
\cite{sarvotham:riedi:baraniuk:2005} studied the dependence structure
of quantities such as session size, duration and transfer rate. They
concluded that it is useful to split the data into two groups
according to the values of peak rate and consider the properties of each group.  These two groups were called alpha
sessions consisting of sessions whose peak rate is above a high
quantile, and beta sessions, 
comprising the remaining traffic. Various criteria for segmenting into
the two groups were considered but always the alpha group was thought of as
sessions corresponding to ``power users" who transmit large files
at large bandwidth, and the beta group
was the remaining sessions. This analysis yielded the following: 
\begin{itemize}
\item A tiny alpha group relative to a huge beta group. In addition,
  it appeared that the alpha group was the major source of burstiness. 
\item A dependence structure that is quite different in the alpha and
  beta groups, with approximate independence between rate and size for the alpha
  group and approximate independence between rate and duration for the beta
  group. To see this, \cite{sarvotham:riedi:baraniuk:2005} measured
  dependence with correlations between the log-variables. 
\end{itemize}

We wondered if the large beta group should be treated as one homogeneous collection of users, especially when one is happy to identify a small and distinct alpha group. Thus, we have investigated whether segmenting the beta group further produces meaningful information.

Section \ref{sec:defns} contains more details on the network traffic
traces that we study, and gives the precise definition of session, size, duration,
rate and peak rate. Historically \citep{crovella:bestavros:1997,
leland:taqqu:willinger:wilson:1994,
willinger:taqqu:leland:wilson:1995a,
willinger:taqqu:sherman:wilson:1997}, data collection was over
finely resolved time intervals, and thus a natural definition of peak
rate is based on computing the maximum transfer rate over consecutive
time slots. We discuss in Section \ref{sec:defns} that this definition
may be flawed due to the choice of the time window resolution giving the
peak rate undesired properties. Thus we 
propose our own definition of peak rate. 

In Section \ref{sec:marg}, we study the marginal distributions of size, duration and rate, and in Sections \ref{sec:depev} and \ref{sec:depcev} we explore the dependence structure between these three variables. Throughout these sections, we depart from the approach of \citet{sarvotham:riedi:baraniuk:2005} by not just looking at the alpha and beta groups; instead, we have split the data into $q$ groups of approximately equal size according to the quantiles of peak rate. Thus, where we previously had a beta group, we now have $q-1$ groups, whose peak rates are in a fixed quantile range. We show that the alpha/beta split is masking further structure and that it is important to take into account the explicit level of the peak rate. In Sections \ref{sec:depev} and \ref{sec:depcev}, we also review and use methods that are more suitable than correlation in the context of heavy tailed-modeling for studying the dependence of two variables.

We also have considered in Section \ref{sec:poiss} whether 
session starting times can be described by a Poisson process. While
several authors have shown that the process of packet arrivals to
servers cannot be modeled under the framework of Poisson processes
\citep{paxson:floyd:1995, willinger:taqqu:sherman:wilson:1997,
willinger:paxson:1998, hohn:veitch:abry:2003}, some argue that the network traffic is
driven by independent human activity and thus justify the search for
this underlying Poisson structure at higher levels of aggregation
\citep{park:shen:marron:hernandez-campos:veitch:2006}. We have found
that despite its inadequateness to describe the overall network
traffic, a homogenous Poisson process is a good model for the session
initiation times within each of the $q$ groups produced by our
segmentation of the overall traffic. In Section \ref{sec:final} we
conclude with some final thoughts including a rough outline for
simulation of data sets based on the aforementioned Poisson framework,
and give possible lines of future study.

\section{Definitions}\label{sec:defns}
\subsection{Size $S$, duration $D$, and rate $R$ of e2e sessions}\label{sub:defns}
Transmissions over a packet-switched computer network do not take
place in a single piece, but rather in several small packets of data
of bounded maximum size that depends on the specific network
protocol. Thus, packet-level network traffic traces consist of records
of packet headers, containing information of each individual packet
such as arrival times to servers, number of bytes transmitted, source
and destination network addresses, port numbers, network protocols,
etc. As the packets travel across the network, 
 routers and switches use the packet header information to move
each packet to its correct destination. The two main goals of
packet-switching are to optimize the utilization of available line
bandwidth and to increase the robustness of communication \citep[see e.g.][]{keshav:1997}. 

The nature of the network data sets presents a challenge for
modeling user behavior. Models such as a superimposition of on-off
processes \citep{willinger:taqqu:sherman:wilson:1997,
sarvotham:riedi:baraniuk:2005} or an infinite-source Poisson
model \citep{guerin:nyberg:perrin:resnick:rootzen:starica:2003,
maulik:resnick:rootzen:2002, dauria:resnick:2006, 
dauria:resnick:2008} require a way to reconstruct  from the
individual packets either a suitable on-period for the former model,
or a suitable transmission session, for the latter. One possible
approach \citep{sarvotham:riedi:baraniuk:2005,
willinger:taqqu:sherman:wilson:1997} is to define an
\textit{end to end (e2e) session}, or briefly session, as a cluster of
bytes with the same source and destination network addresses, such
that the delay between any two successive packets in the cluster is
less than a threshold $t$. A session plays the role of an arriving
entity in an infinite-source Poisson model or the role of an on-period
in an on-off process model. 

For each session, we have the following variables:
\begin{itemize}
\item $S$ represents the size, that is, the number of bytes transmitted.
\item $D$ represents the duration, computed as the difference in seconds between the arrival times of the first and last packets in the session.
\item $R$ represents the average transfer rate, namely $ S/D$.
\end{itemize}
	
Note that $R$ is not defined for single-packet sessions, for which $D$
by definition is zero. More generally, sessions with very small $D$
may also be problematic to handle. 
For instance,
 it would be hard to
  believe that a session sending only two packets back-to-back  has an
  $R$ that equals the line bandwidth. In order to avoid this issue,
  for our analysis we ignore sessions with $D<100 ms$.
See \cite{zhang:breslau:paxson:shenker:2002} for related  comments.

\subsection{Predictors of burstiness}\label{subsec:predburst}

In addition to $S$, $D$ and $R$, \cite{sarvotham:riedi:baraniuk:2005}
consider a fourth quantity which serves  as an explanatory variable
for burstiness, namely the session's maximum input in consecutive time
windows. A closely related variable arises by considering the
session's peak rate in consecutive intervals. In what follows, we
review the properties of these two variables and show that they are
not ideal for describing burstiness. Therefore, we will propose a different
definition of peak rate.

\subsubsection{The $\delta$-maximum input.}\label{subsubsec:deltamax}

\begin{figure}
\includegraphics{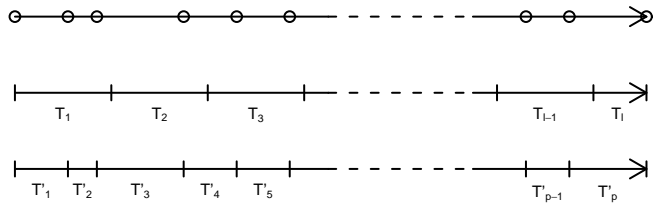}
\caption{\textit{Top arrow} Representation of a typical session; here each packet is depicted as an oval \textit{Middle arrow} \cite{sarvotham:riedi:baraniuk:2005}'s division approach \textit{Bottom arrow} Our proposed division according to the packet arrival times}\label{fig:session}
\end{figure}

Fix a small $\delta >0$ and divide each session in $l$ subintervals of
length $\delta$, where $l=\lceil D/\delta\rceil$ (see Fig. \ref{fig:session}, Top and Middle).
For $i=1,\ldots,l$,  define the following auxiliary variables:
\begin{itemize}
\item $B_i$ represents the number of bytes transmitted over the $i$th
  subinterval of the session.
\item $T_i$ represents the duration of the $i$th subinterval. For
  $i=1,\ldots,l-1,$ we have $T_i=\delta$. However, notice that
  $T_l=D-(l-1)\delta$. 
\end{itemize}

The $\delta$-\textit{maximum input} of the session is defined as
$I_{\delta}=\bigvee_{i=1}^n B_i.$ This $I_{\delta}$ is the original variable used by
\cite{sarvotham:riedi:baraniuk:2005}.

\subsubsection{The $\delta$-peak rate.}\label{subsubsec:deltapeak}
If the goal is to
explain burstiness, a natural alternative to maximum input is to
consider rates
in consecutive time subintervals, rather
than inputs. This yields a closely 
related predictor: the $\delta$-peak rate. 
The definition of the $\delta$-\textit{peak rate} for a session, denoted as
$R_{\delta}$, relies on the \cite{sarvotham:riedi:baraniuk:2005}'s division of the session (see Fig. \ref{fig:session}). 
We define $R_{\delta}=\bigvee_{i=1}^n
{B_i}/{T_i}$. 

Observe the following properties for a session:
\begin{enumerate}[(i)]
\item $\sum_{i=1}^n B_i = S$;
\item $\sum_{i=1}^n T_i=D$;
\item $R_{\delta}\geq R$. To see this, note
\begin{align*}
R= \frac SD =\frac{\sum_{i=1}^n T_i \cdot
  \frac {B_i}{T_i}}{\sum_{i=1}^n T_i}
\leq \bigvee_{i=1}^n B_i/T_i =R_\delta.
\end{align*}
\end{enumerate}

A quick analysis shows that the last property does not necessarily hold if we do not carefully 
define the duration of the last subinterval $T_n$ as above, but
instead  set $T_n=\delta$. For a numerical example, let $\delta=1$ and
consider a session with $n=2, B_1=B_2=1,D=1.1$. Using the wrong
definition $T_n=\delta$ yields $T_1=T_2=1$, hence the average transfer
rate $R=(B_1+B_2)/D=2/1.1$ but the peak transfer rate
$R_\delta=\max\left\{{B_1}/{T_1},{B_2}/{T_2}\right\}=1<2/(1.1)$. 

While both $I_\delta$ and $R_\delta$ appear to be natural predictors of
burstiness, they both possess undesirable properties. They both depend on
the parameter $\delta$
which is not an intrinsic characteristic of the session.
As $\delta\downarrow0$, many
consecutive subintervals thus
have a single packet, as in Fig. \ref{fig:session}.
Therefore, as $\delta \downarrow 0$,
\begin{itemize}
\item $I_\delta\rightarrow$\emph{ maximum packet size}, 
which precludes  $I_\delta$ from being a
 useful measure of burstiness.

\item $R_\delta\rightarrow\infty$, implying that $R_\delta$ is much
  greater than the line capacity as $\delta\rightarrow0$. In fact, for
  this limit to hold, it is sufficient that a single subinterval has a
  few packets, which suggests that the convergence rate of
  $R_\delta\rightarrow\infty$ is greater than the convergence rate of
  $I_\delta\rightarrow$\emph{maximum packet size}. This implies that
  we still may have unreasonably large $R_\delta$s for relatively
  large $\delta$s. Therefore, the interpretation of $R_\delta$ as an
  actual peak transfer rate becomes problematic.
\end{itemize}

Owing to the drawbacks of the previous two definitions,
we propose our own definition of peak rate.

\subsubsection{Peak rate $R^\vee$.}\label{subsubsec:truePeakRate}
Suppose a session has $p$ packets (see Fig. \ref{fig:session}, Bottom). 
Consider the following variables.

\begin{itemize}
\item $B'_i$ represents the number of bytes of the $i$th packet.
\item $T'_i$ represents the interarrival time of the $i$th and $(i+1)$th packets, $i=1,\ldots,p-1$.
\end{itemize}

For $k=2,\ldots,p,$ we define the \textit{peak rate of order $k$}, denoted by $R^{(k)}$, as
\begin{equation}\label{pratek}
R^{(k)}=\bigvee_{j=1}^{p-k+1} \frac {\sum_{i=j}^{j+k-1} B'_i} {\sum_{i=j}^{j+k-2} T'_i}.
\end{equation}

In the above definition, the quotient measures the actual transfer rate of a stream of bytes consisting of $k$ consecutive packets. For a session consisting of $p$ packets, there are $p-k+1$ streams of $k$ consecutive packets, hence $R^{(k)}$ is a measure of the actual peak transfer rate when only $k$ consecutive packets are taken into account.
We then define the \textit{peak rate} as
\begin{equation}\label{prate}
R^\vee=\bigvee_{k=2}^p R^{(k)}.
\end{equation}
Notice that $R^\vee\geq R^{(p)}=R$.

As opposed to  $I_{\delta}$ and $R_{\delta}$, $R^\vee$ does not depend
on  external parameters such as $\delta$, and thus it is an 
 intrinsic
characteristic of a session. In addition, $R^\vee$ inherits the interpretation of
$R^{(k)}$ and therefore it may be interpreted itself as a measure of
the actual maximum transfer rate taken over all possible streams of
consecutive packets. A drawback of our definition is that $R^\vee$ is
complex to analyze mathematically.

\subsection{The data set}\label{subsec:dataset}

We present our results for a network trace captured at the University
of Auckland between December 7 and 8, 1999, which was publicly
available as of May 2009 through the National Laboratory for Applied
Network Research website at
\url{http://pma.nlanr.net/Special/index.html}.
 Auckland's data set is a collection of GPS-synchronized traces, where all non-IP traffic has been discarded and only TCP, UDP and ICMP traffic is present in the trace. We have taken the part of the trace corresponding exclusively to incoming TCP traffic sent on December 8, 1999, between 3 and 4 p.m. We have found that our results hold for several other data sets. See Section \ref{sec:final} for more details about this and other data traces.

The raw data consists of 1,177,497 packet headers, from which we
construct 44,136 sessions using a threshold between sessions of $t=2s$
and considering only those sessions with $D>100ms$ (as explained in
the last paragraph of Section \ref{sub:defns}). We have found similar
results for various choices of thresholds between sessions, including
$t=0.1, 0.5, 10, 60, 100s$, but here we only present our results for $t=2s$.

In addition, for each session we have peak rate $R_i^\vee$ and starting time $\Gamma_i$. Thus, the data set has the form $\{((S_i,D_i,R_i),R_i^\vee,\Gamma_i);1\leq i\leq44,136\}$, that is, a set of 5-tuples, but with the above notation we emphasize that the primary interest is placed on the dependence structure of triplet $(S_i,D_i,R_i)$.

We split these sessions into $10$ groups of approximately equal size
according to the empirical deciles of $R^\vee$. Thus,  all the
sessions in the $g$th  group, $g=1,\dots, 10$, have $R^\vee$ is in a
fixed decile range, $(10(g-1)\%,10g\%]$.
Hence we term the group of sessions ``the $g$th decile group",
$g=1,\ldots,10$. Therefore, where \cite{sarvotham:riedi:baraniuk:2005} had alpha
and beta groups, we now have a more refined segmentation.

In the remainder of the paper, we show that this
refined split reveals features that are hidden by an elementary
alpha/beta split.

\section{Marginal distributions of $S$, $D$ and $R$.}\label{sec:marg}

In what follows, we analyze the marginal distributions of $S$, $D$ and
$R$ in the $10$ different decile groups to check for the presence of
one stylized fact in network data sets, namely, heavy tails. We found
that $S$ and $D$ have heavy tails for all the different decile groups,
but not $R$. Let us start by discussing background.

\subsection{Heavy tails and maximal domains of attraction}\label{subsec:mda}

A positive random variable $Y$ has \textit{heavy tails} if its
distribution function $F$ satisfies 
\begin{equation}\label{eq:heavytail}
1-F(y)=\bar{F}(y)=y^{-1/\gamma}L(y),
\end{equation}
where $L$ is a slowly varying function and $\gamma>0$. We also say that $F$ is heavy tailed and we call $\gamma$ the shape parameter.
When $F$ satisfies Eq. \ref{eq:heavytail}, it is also said to have regularly varying tails with tail index $1/\gamma$. Equation \ref{eq:heavytail} is equivalent to the existence of a sequence $b_n\rightarrow\infty$ such that
\begin{equation}\label{eq:regvar}
\mu_n(\cdot):=n\mathbb{P}\left[\frac{Y}{b_n}\in\cdot\right]\xrightarrow{v}c\nu_{\gamma}(\cdot),
\end{equation}
vaguely in $M_+(0,\infty]$, the space of Radon measures on
$(0,\infty]$. Here $\nu_\gamma(x,\infty]=x^{-1/\gamma}$ and
$c>0$. Equation \ref{eq:regvar} is important for
generalizing the concept of heavy tailed distributions to higher
dimensions.

An important concept is
maximal domains of attraction. Suppose $\{Y_i;i\geq1\}$ is iid with common distribution $F$. The distribution $F$ is in the \textit{maximal domain of attraction} of the extreme value distribution $G_{\gamma}$, denoted  $F\in\mathcal{D}(G_{\gamma})$, if there exist sequences $a_n>0$ and $b_n\in\mathbb{R}$ such that for $y\in\mathbb{E}^{(\gamma)}=\{y\in\mathbb{R}:1+\gamma y>0\}$:
\begin{equation}\label{eq:mda}
\lim_{n\rightarrow\infty}\mathbb{P}\left[\frac{\bigvee_{i=1}^nY_i-b_n}{a_n}\leq y\right]=G_{\gamma}(y):=\exp{\{-\left(1+\gamma y\right)^{-1/\gamma}\}}.
\end{equation}
This is equivalent to the existence of functions $a(t)>0$ and $b(t)\in\mathbb{R}$ such that for $y\in\mathbb{E}^{(\gamma)}$:
\begin{equation}\label{eq:mdab}
\lim_{t\rightarrow\infty}t\mathbb{P}\left[Y_1>a(t)y+b(t)\right]=-\log G_{\gamma}(y).
\end{equation}
The class of distributions $\mathcal{D}(G_{\gamma})$ is known as the Fr\'echet domain when $\gamma>0$, Gumbel domain when $\gamma=0$ and Weibull domain when $\gamma<0$. Equation \ref{eq:mda} is also known as the extreme value condition.

For $\gamma>0$, 
\begin{equation}
\bar{F}(y)=y^{-1/\gamma}L(y) \Leftrightarrow F\in\mathcal{D}(G_{\gamma}),
\end{equation}
for some slowly varying $L$. In other words, a necessary and
sufficient condition for a distribution to be heavy tailed is that it
is in the Fr\'echet class \citep{dehaan:ferreira:2006,
resnick:1987}.

Thus, the focus will be placed on checking the Fr\'echet domain condition. First, we check whether or not the marginals of $(S,D,R)$ are in some domain of attraction. If that turns out to be the case, we proceed to check for the specific domain class.

\subsection{Domain of attraction diagnostics}\label{sub:diag}

\subsubsection{Excesses over high thresholds}\label{sub:excess}

One common method \citep{davison:smith:1990,
beirlant:goegebeur:teugels:segers:2004,coles:2001,
reiss:thomas:2007,mcneil:frey:embrechts:2005,
dehaan:ferreira:2006} to check the extreme value condition, given by
Eq. \ref{eq:mda}, relies on threshold excesses, using all data that are
``extreme'' in the sense that they exceed a particular designated high
level. 

More precisely, consider a random variable $Y$ with distribution function $F$. Given realizations of $Y$, say $y_1,\ldots,y_n$ and a threshold $u$, we call $y_j$ an exceedance over $u$ if $y_j>u$, and in such case, $y_j-u$ is called the \textit{excess}. Denote the \textit{excess distribution} over the threshold $u$ as $F_u$, i.e.
\begin{equation*}
F_u(y)=\mathbb{P}[Y-u\leq y|Y>u],
\end{equation*}
for all $0\leq y\leq y_F-u$, where $y_F\leq\infty$ is the right endpoint of $F$.
The connection with domains of attraction is that
\begin{equation}\label{eq:excess}
F\in\mathcal{D}(G_{\gamma}) \Leftrightarrow  \lim_{u \to y_F}\sup_{0\leq y\leq y_F-u}\left|F_u(y)-GPD_{\gamma,\beta(u)}(y)\right|=0,\textrm{ for some $\beta(u)>0$}.
\end{equation}
Here $GPD_{\gamma,\beta}$, with $\gamma\in\mathbb{R},\beta>0$ is the generalized Pareto distribution , defined as
\begin{equation*}
GPD_{\gamma,\beta}(y):=1-\left(1+\gamma y/\beta\right)^{-1/\gamma},
\end{equation*}
for $y\geq0$ when $\gamma\geq0$ and $0\leq y\leq -\beta/\gamma$ when
$\gamma<0$. See  \cite{pickands:1975,balkema:dehaan:1974,dehaan:ferreira:2006}.

For a distribution $F$, the method of excesses over high thresholds
(also referred to as peaks over thresholds or {\it POT\/}) 
assumes equality in  Eq. \ref{eq:excess} holds for a high threshold
$u$, without need
to take a limit, meaning that the excess distribution over such $u$
equals a generalized Pareto distribution. See
\cite{embrechts:kluppelberg:mikosch:1997, coles:2001,
reiss:thomas:2007,
dehaan:ferreira:2006}.
Thus, suppose
$Y_1,\ldots,Y_n$ are iid with common distribution $F$ and let
$Y_{1:n}\leq Y_{2:n}\leq\cdots\leq Y_{n:n}$ be the order
statistics. Fix a high threshold $\hat{u}=Y_{n-k:n}$ as the $(k+1)$th
largest statistic, and fit a $GPD_{\gamma,\beta}$ model to
$Y_{n-k+1:n}-\hat{u},\ldots,Y_{n:n}-\hat{u}$. Then the evidence
supports $F\in\mathcal{D}(G_{\gamma})$ if and only if for some high
threshold $\hat{u}$ that fit is adequate. For informally assessing the
goodness of fit, we compare 
via quantile-quantile
(QQ) plots
the sample quantiles, namely
$\hat{Y}_{n-k+1:n}-\hat{u},\ldots,\hat{Y}_{n:n}-\hat{u}$, against the
theoretical quantiles given by the $GPD$ fit. It is not difficult to show that $Z\sim
GPD_{\gamma,\beta}$ is equivalent to the statement that
$\log\left(1+\gamma Z/\beta)\right)/\gamma\sim\exp(1)$, and so we draw
QQ plots in this latter scale after estimating $\gamma,\beta$ by means
 of, say, maximum likelihood.
 
\begin{figure}[htb]
\includegraphics[scale=1.2]{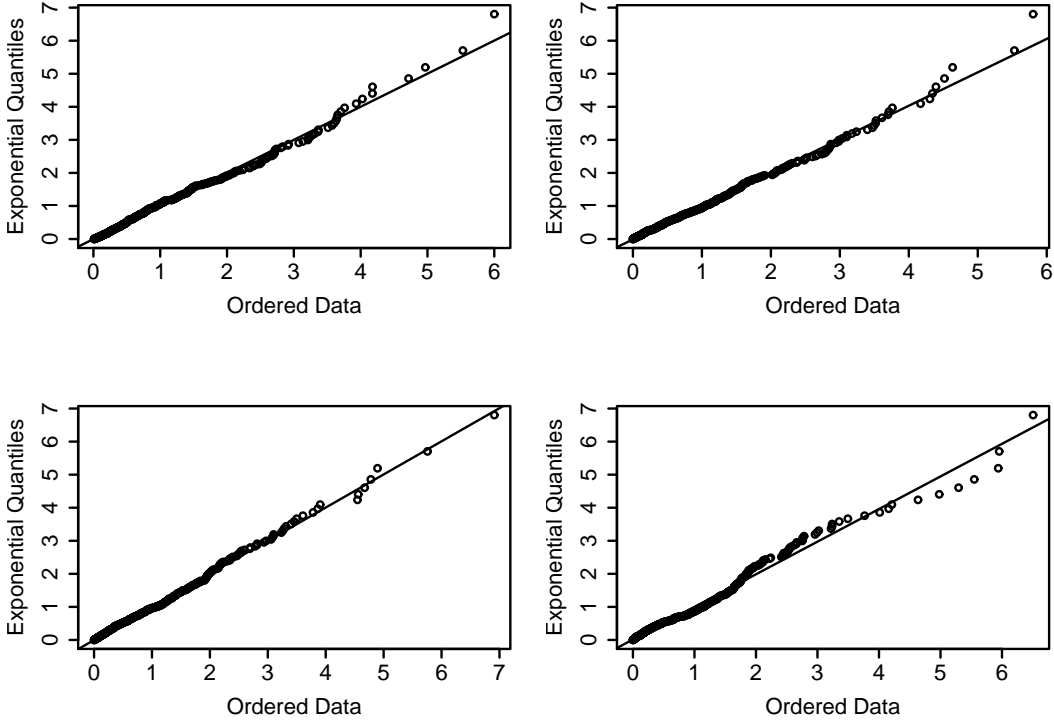}
\caption{$GPD$ QQ-plots of excesses; a number $k=450$ of upper order statistics is used for each fit \textit{Upper left} Size in the 10th decile group \textit{Upper right} Duration in the 10th decile group \textit{Lower left} Rate in the 10th decile group \textit{Lower right} Rate in the 4th decile group}\label{fig:gpdqq}
\end{figure}

Using the POT method, 
 we found no evidence against $F_S, F_D\in\mathcal{D}(G_{\gamma})$ for
 all the 10 decile groups. Typical QQ plots are those corresponding to
 the $GPD_{\gamma,\beta}$ fit for the excess of $S$ and $D$ in the
 10th decile group, shown in Fig. \ref{fig:gpdqq}, \textit{Upper left} and \textit{Upper right} panels, respectively. Both plots exhibit an almost
 perfect straight line. We also found that the QQ plots corresponding
 to the excesses of $S$ and $D$ in all the other decile groups exhibit
 straight line trends, showing thus no evidence against satisfaction
 of the extreme
 value condition. 

Similarly, Fig. \ref{fig:gpdqq} \textit{Lower left} panel exhibits the QQ plot of the $GPD_{\gamma,\beta}$ fit for the excess of $R$ in the 10th decile group, which shows no evidence against $F_R\in\mathcal{D}(G_{\gamma})$. However, for all the other decile groups, we found evidence against $F_R\in\mathcal{D}(G_{\gamma})$. For instance, a QQ plot of the $GPD_{\gamma,\beta}$ fit for the excess of $R$ in the 4th decile group is shown in Fig. \ref{fig:gpdqq} \textit{Lower right} panel, exhibiting a major departure from the straight line. We also found no straight line trend in the rest of the QQ plots of the $GPD_{\gamma,\beta}$ fit for the excess of $R$ in the lower nine decile groups.

\subsubsection{Formal tests of domain of attraction}\label{sub:formaltest}

Recently, two formal methods for testing $F\in\mathcal{D}(G_{\gamma})$
have been derived by \cite{dietrich:dehaan:husler:2002} and
\cite{drees:haan:li:2006}. Both tests are similar in that the two are
based on quantile function versions of the well known Cr\'amer
von-Mises and Anderson-Darling test statistics \citep[see e.g.][]{lehmann:romano:2005}, respectively, for checking the goodness
of fit of a given distribution. In addition, both tests assume a so
called second order condition which is difficult to check in
practice. A follow-up study by \cite{husler:li:2006} examines the two
tests' error and power by simulations. A thorough discussion of these
tests and the second order condition is provided by
\cite{dehaan:ferreira:2006}. Here we review the method proposed in
\cite{dietrich:dehaan:husler:2002} and use it to test for the
extreme value condition. 

\cite{dietrich:dehaan:husler:2002} state that if
$F\in\mathcal{D}(G_{\gamma})$ for some $\gamma\in\mathbb{R}$ and also
if $F$ satisfies an 
additional second order tail condition \citep[for the details of this condition, see][equation (4)]{dietrich:dehaan:husler:2002}, then: 
\begin{align}
E_{k,n}:=& k\int_0^1 \left(\frac{\log Y_{n-[kt]:n}-\log
    Y_{n-k:n}}{\hat{\gamma}_+}
  -\frac{t^{-\hat{\gamma}_-}-1}{\hat{\gamma}_-}\right)t^2 dt \nonumber
\\
\xrightarrow{d} &E_{\gamma}:=\int_0^1 \Bigl( (1-\gamma_-)(t^{-\gamma_-
    -1}W(t)-W(1)) - (1-\gamma_-)^2
  \frac{t^{-\gamma_-}-1}{\gamma_-}P_{\gamma_-} \nonumber \\
{}&\qquad
 +\frac{t^{-\gamma_-}-1}{\gamma_-}R_{\gamma_-} + (1-\gamma_-)R_{\gamma_-}\int_t^1 s^{-\gamma_- -1}\log sds \Bigr)^2t^2dt,\label{eq:dietrich}
\end{align}
as $k\rightarrow\infty,k/n\rightarrow0,n\rightarrow\infty$ and
$k^{1/2}A(n/k)\rightarrow0$, where $A$ is related to the second order
condition, $\gamma_+=\max\{\gamma,0\}$ and
$\gamma_-=\min\{\gamma,0\}$, $W$ is a Brownian motion, $P_{\gamma_-}$
and $R_{\gamma_-}$ are some integrals involving $W$ \citep[for the details,
see][]{dietrich:dehaan:husler:2002,dehaan:ferreira:2006},
and $\hat{\gamma}_+$ and $\hat{\gamma}_-$ are consistent estimators of
the corresponding parameters.

In practice, \cite{dietrich:dehaan:husler:2002} recommend replacing $\gamma$ by its estimate. Therefore, based on Eq. \ref{eq:dietrich}, we could test
$$H_0:F\in\mathcal{D}(G_\gamma),\gamma\in\mathbb{R}+ \text{second
  order condition}$$
 by first determining the corresponding quantile
$Q_{1-\alpha,\hat{\gamma}}$ of the distribution $E_{\hat{\gamma}}$ and
then comparing it with the value of $E_{k,n}$. If
$E_{k,n}>Q_{1-\alpha,\hat{\gamma}}$ we reject $H_0$ with asymptotic
type I error $\alpha$ and otherwise there is no evidence to reject
$H_0$. Notice that this is a one-sided test of hypothesis, but a
two-sided test could be performed in a similar fashion. 

A drawback of this test is that we must include in $H_0$ the
additional second order condition, which is difficult to check in
practice. While many common distributions satisfy the second order
condition, including the normal, stable, Cauchy, log-Gamma, among others,
the Pareto distribution is a notable example of a distribution which
does not satisfy the second order condition.

In addition, there are two other drawbacks of this test. First, it is
based on the usual setting of acceptance-rejection regions, and thus
it provides no measure of the strength of rejection of $H_0$. While
this typically is addressed with the equivalent setting based on
p-values, the limit distribution in Eq. \ref{eq:dietrich} is
analytically intractable and so are the p-values. Second, since the
limit in Eq. \ref{eq:dietrich} depends on $k$, the conclusions of the test
are also highly dependent on the choice of $k$.

\cite{dietrich:dehaan:husler:2002} state a corollary in which the
limit distribution in Eq. \ref{eq:dietrich} is greatly simplified by
observing that $\gamma_-=0$ for all $\gamma\geq0$. This result
is easier to apply. Under the assumption that $F\in\mathcal{D}(G_\gamma),\gamma\geq0$ and the second order condition, Eq. \ref{eq:dietrich} becomes:
\begin{align}
\tilde{E}_{k,n}=&k\int_0^1\left(\frac{\log Y_{n-[kt]:n}-\log
    Y_{n-k:n}}{\hat{\gamma}_{k,n}}+\log t\right)^2t^2dt\nonumber \\
\xrightarrow{d}&\tilde{E}=\int_0^1\left(t^{-1}W_t-W_1+\log
  t\int_0^1(s^{-1}W_s-W_1)ds\right)^2t^2dt.\label{eq:pdietrich}
\end{align}

Suppose $\tilde{E}_1,\ldots,\tilde{E}_N$ is a random sample of $\tilde{E}$, that we can
obtain by simulation since the limit distribution in
Eq. \ref{eq:pdietrich} is free of unknown parameters. Based on
Eq. \ref{eq:pdietrich}, we propose the following test for 
$$H_0:F\in\mathcal{D}(G_\gamma),\gamma\geq0+ \text{second order
  condition}.$$ Estimate a (one-sided) p-value
$p(k)=\mathbb{P}(\tilde{E}>\tilde{E}_{k,n})$ as the relative frequency 
$$\hat{p}(k)=\frac 1N \sum_{j=1}^N 1_{\tilde{E}_{j}>\tilde{E}_{k,n}}.$$ If
$\hat{p}(k)<\alpha$, then reject $H_0$ with an asymptotic type I error
$\alpha$, otherwise there is no evidence to reject $H_0$. With this
method, the p-values give a measure of the strength of rejection of
$H_0$. Furthermore, we can check the stability of the conclusion of
the test as a function of $k$ by constructing the plot
$\{(k,\hat{p}(k));k\textrm{ in an appropriate range}\}$. The range of
values of $k$ is chosen to accommodate for the limit in
Eq. \ref{eq:pdietrich}, namely
$k\rightarrow\infty,k/n\rightarrow0,n\rightarrow\infty$. For example,
\cite{husler:li:2006} found via simulations that the power of the test
in 
\cite{dietrich:dehaan:husler:2002}  appears to be high for $k$ such
that $k/n\approx0.05$, at least for their various choices of $F$. To
compute $\tilde{E}_{k,n}$, we use $\hat{\gamma}_{k,n}$ given by the consistent
Hill estimator \citep{hill:1975} or perhaps maximum likelihood if we
suspect $\gamma =0$.

\begin{figure}[tb]
\includegraphics{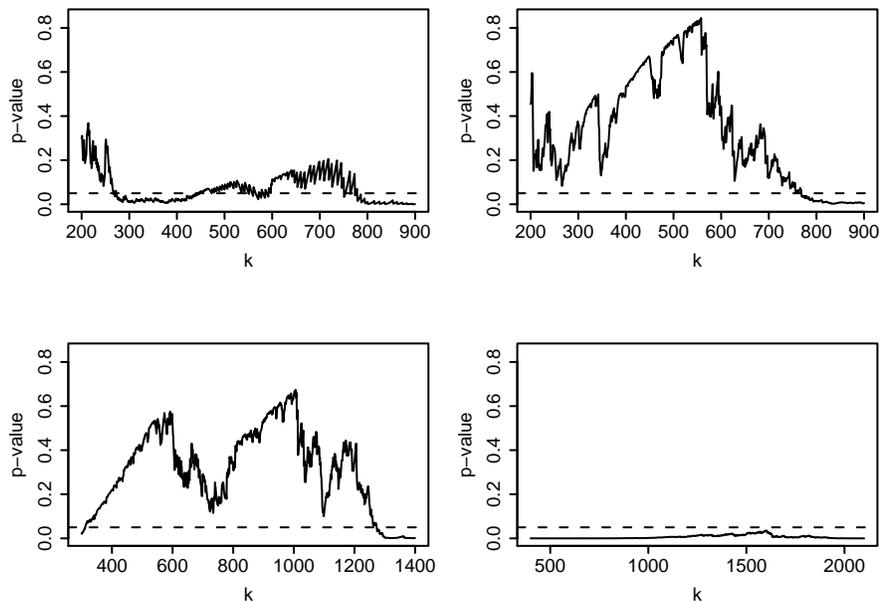}
\caption{Plots of p-values as a funtion of $k$ for the test of the extreme value condition for the distribution of the following variables; a horizontal dashed line is drawn at $\alpha=0.05$ \textit{Upper left} Size in the 10th decile group \textit{Upper right} Duration in the 10th decile group \textit{Lower left} Rate in the 10th decile group \textit{Lower right} Rate in the 4th decile group}\label{fig:pvalues}
\end{figure}

We use this method with an asymptotic
nominal type I error $\alpha=0.05$. We found no evidence against $F_S,
F_D\in\mathcal{D}(G_{\gamma}),\gamma\geq0$ for all the 10 decile
groups. Typical plots of the p-values $\hat{p}(k)$ for the variables
$S$ and $D$ are those corresponding to the 10th decile group, shown in
Fig. \ref{fig:pvalues} \textit{Upper left} and \textit{Upper right} panels,
respectively. Both plots exhibit that $\hat{p}(k)>0.05$ for a wide
range of values of $k$. We also found here that the plots
$\{(k,\hat{p}(k))\}$ corresponding to all the other decile groups show
no evidence against $F_S, F_D\in\mathcal{D}(G_{\gamma}),\gamma\geq0$. 
Coupled with the evidence from the QQ plots, we believe $\gamma>0.$

Similarly, Fig. \ref{fig:pvalues} \textit{Lower left} exhibits the plot
$\{(k,\hat{p}(k))\}$ corresponding to the distribution of $R$ in the
10th decile group. Once again we found that $\hat{p}(k)>\alpha$ for a
wide range of values of $k$, thus showing  no evidence against 
$F_R\in\mathcal{D}(G_{\gamma}),\gamma\geq0$. However, we did find
evidence against $H_0:F_R\in\mathcal{D}(G_\gamma),\gamma\geq0+
\textit{second order condition}$ for all the lower nine decile
groups. A typical example of the plot $\{(k,\hat{p}(k))\}$ in these
latter groups is exhibited in Fig. \ref{fig:pvalues} \textit{Lower right} for the 4th
decile group, which shows that $\hat{p}(k)$ are significantly lower
than 0.05 across a wide range of $k$ values.

Therefore, for the lowest nine decile groups, we reject
$H_0:F_R\in\mathcal{D}(G_\gamma),\gamma\geq0+$\textit{second order condition}. One possible alternative is that $F_R\in\mathcal{D}(G_\gamma),\gamma<0$, or equivalently, that $x_{F_R}<\infty$ and $F_{(x_{F_R}-R)^{-1}}\in\mathcal{D}(G_{-1/\gamma})$ \citep{dehaan:ferreira:2006, resnick:1987}. Hence, by applying the above test to $H_0:F_{(x_{F_R}-R)^{-1}}\in\mathcal{D}(G_\gamma),\gamma\geq0+$\textit{second order condition}, we dropped the possibility of this new $H_0$ because the $\hat{p}(k)<0.05$ for a wide range of values of $k$ for the lower nine decile groups. Here we estimated $x_{F_R}$ with $R_{n:n}+1/n'$ for a high value of $n'$. 

This last result left us with two possibilities. Either $F_R\not\in\mathcal{D}(G_\gamma),\gamma\in\mathbb{R}$ or simply, the additional second order condition does not apply for $F_R$ (and in
this case we still may have $F_R\in\mathcal{D}(G_\gamma),\gamma\geq0$). As previously mentioned, the second order condition is difficult to check in practice without any prior knowledge of the distribution function, and thus the hypothesis test fails to provide a clear description of the distribution $F_R$. Nevertheless, in Section \ref{sec:depcev} we are able to say something about $F_R$.

\subsection{Estimation}\label{subsec:estimation}

In Section \ref{sub:formaltest} we showed that $F_S,F_D\in\mathcal{D}(G_\gamma),\gamma\geq0$ for all the decile groups and $F_R\in\mathcal{D}(G_\gamma),\gamma\geq0$ only for the 10th decile group. We now proceed to the estimation of the shape parameter $\gamma$ for these distributions.

The Hill estimator is a popular estimator of $\gamma$ \citep{hill:1975,csorgo:deheuvels:mason:1985,davis:resnick:1984,dehaan:resnick:1998,hall:1982}. The \textit{Hill estimator} based on the $k$ largest order statistics is
\begin{equation}\label{eq:hill}
\hat{\gamma}_{k,n}=\frac 1k \sum_{i=n-k+1}^n \log\frac{Y_{i:n}}{Y_{n-k:n}},\quad k=1,\ldots,n-1.
\end{equation}

For $F\in\mathcal{D}(G_\gamma),\gamma>0$, the Hill estimator $\hat{\gamma}_{k,n}$ is a consistent estimator of $\gamma$. Furthermore, under an additional second order condition of the type needed in Eq. \ref{eq:dietrich}:
\begin{equation}\label{eq:hillnorm}
\sqrt{k}(\hat{\gamma}_{k,n}-\gamma)\xrightarrow{d} N(0,\gamma^2),
\end{equation}
so both consistency and asymptotic normality hold as
$k\rightarrow\infty,k/n\rightarrow0,$ and $n\rightarrow\infty$. See,
for example,  \cite{dehaan:ferreira:2006, resnickbook:2007,
geluk:dehaan:resnick:starica:1997,
dehaan:resnick:1998,
peng:1998,
dehaan:peng:1998, mason:turova:1994}.

\begin{figure}[tb]
\includegraphics{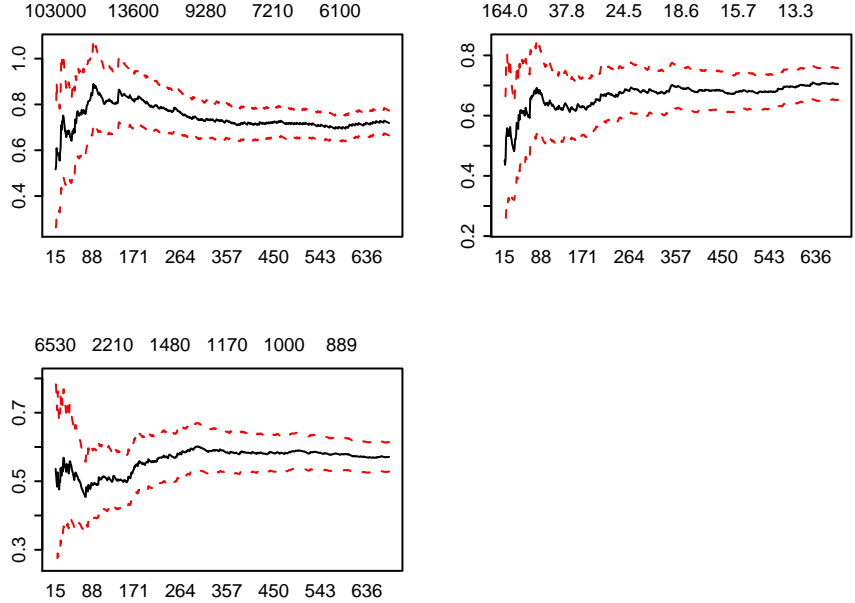}
\caption{Hill plots, with 95\% confidence bands in dashed lines, corresponding to the shape parameter $\gamma$ of the variables in the 10th decile group; the values on top give thresholds and the values on bottom indicate the number of upper order statistics \textit{Upper left} Size \textit{Upper right} Duration \textit{Lower left} Rate}\label{fig:hillplots}
\end{figure}

The Hill estimator depends on the number $k$ of upper-order statistics and so in practice, we make a \textit{Hill plot} $\{(k,\hat{\gamma}_{k,n});k\geq1\}$ and pick a value of $\hat{\gamma}_{k,n}$ for which the graph looks stable. Figure $\ref{fig:hillplots}$ exhibits Hill plots for the shape parameter $\gamma$ of the distribution of $S$, $D$ and $R$ for the 10th decile group. The three plots show stable regimes for $\gamma$ around $k=450$. We also found stability in the Hill plots for the shape parameter $\gamma$ of the distribution of $S$, $D$ in all the other decile groups.

Table \ref{tab:hill} contains the Hill estimates of $\gamma$ for our
data set, along with estimates of the asymptotic standard error based
on Eq. \ref{eq:hillnorm}. A number $k$ of upper order statistics was
chosen individually for each variable and each decile group based on
the corresponding Hill plots. For most decile groups, we used
$k\approx 400$  $(k/n\approx0.05)$, as suggested by the empirical study by \cite{husler:li:2006}. Notice that the majority of the estimates are greater than 0.5, which implies that the corresponding distributions have infinite variances.

\begin{table}
\caption{Summary of Hill estimates with asymptotic standard errors for the shape parameter of $S$, $D$ and $R$}
\begin{tabular}{l c r c r c c r c r c c r c r}
\hline\noalign{\smallskip}
Decile group & & $\gamma_S$ & & s.e. & & & $\gamma_D$ & & s.e. & & & $\gamma_R$ & & s.e.\\\hline\noalign{\smallskip}
1 & & 0.56 & & 0.056 & & & 0.60 & & 0.028 & & & & &\\
2 & & 0.55 & & 0.061 & & & 0.47 & & 0.023 & & & & &\\
3 & & 0.62 & & 0.044 & & & 0.63 & & 0.034 & & & & &\\
4 & & 0.62 & & 0.036 & & & 0.62 & & 0.029 & & & & &\\
5 & & 0.61 & & 0.035 & & & 0.55 & & 0.029 & & & & &\\
6 & & 0.69 & & 0.040 & & & 0.55 & & 0.028 & & & & &\\
7 & & 0.88 & & 0.042 & & & 0.73 & & 0.037 & & & & &\\
8 & & 0.77 & & 0.045 & & & 0.71 & & 0.033 & & & & &\\
9 & & 0.70 & & 0.037 & & & 0.69 & & 0.032 & & & & &\\
10 & & 0.73 & & 0.034 & & & 0.68 & & 0.032 & & & 0.58 & & 0.027\\\hline\noalign{\smallskip}
\end{tabular}
\label{tab:hill}
\end{table}

We make some final comments on our choice of the Hill estimator against the Pickands estimator \citep{pickands:1975,dekkers:dehaan:1989}. In Section \ref{sub:formaltest} we only showed that $\gamma\geq0$ which is a weaker assumption than the requirement of $\gamma>0$ for the Hill estimator. Unlike the Hill estimator, the Pickands estimator is more robust in the sense that it does not need $\gamma>0$. However, the \textit{Pickands plots} proved to be very unstable for our data set.

\section{Dependence structure of $(S,D,R)$ when the three variables have heavy tails}\label{sec:depev}

We now analyze the dependence structure of the triplet $(S,D,R)$
across the $10$ different decile groups. Since $S=DR$, at most two
of the three components in $(S,D,R)$ may be independent. This makes it
reasonable to focus on the analysis of each pair of variables. We
concentrate on the pairs in $(S,D,R)$ with heavy tailed
marginals and  first focus on the dependence structure of $(S,D)$
across the 10 deciles groups. We later study the dependence structure
of both $(R,S)$ and $(R,D)$, but only in the 10th decile group. For
the other decile groups, we found strong evidence suggesting $R$ does
not have heavy tails, and thus we leave this case for Section
\ref{sec:depcev}.  Our finer segmentation into the deciles of $R^\vee$
reveals hidden features in an alpha/beta split, and therefore it is
important to take into account the explicit level of $R^\vee$. 

One way to assess the dependence structure is with sample
cross-correlations. In heavy-tailed modeling, although the sample
correlations may always be computed, there is no guarantee that the
theoretical correlations exist. Recall  Table \ref{tab:hill} shows that most estimates of $\gamma$ for $S$, $D$ and $R$ are greater than 0.5, and thus correlations do not exist in these instances.
Moreover, correlation is a crude summary of  dependence that is most
informative between jointly normal variables, and that certainly does
not distinguish between the dependence between large values and the
dependence between small values. In the context of data networks, the
likelihood of various simultaneous large values of $(S,D,R)$ may be
important for understanding burstiness. For example, if large values of $D$ are likely to occur simultaneously with large values of $R$, then we can expect a network that is prone to congestion. In this situation, a scatterplot $\{(D_i,R_i)\}$ would be mostly concentrated in the interior of the first quadrant of $\mathbb{R}^2$. On the other hand, if large values of one variable are not likely to occur with large values of the other one, the same scatterplot would be mostly concentrated on the axes.

Understanding network behavior requires a description of the extremal
dependence of $S,D$ and $R$ and this extremal dependence is
conveniently summarized by the spectral measure.
 See \cite{dehaan:resnick:1977,
  dehaan:ferreira:2006, resnickbook:2007, resnickbook:2008}.
We begin by
discussing  important concepts.

\subsection{Bivariate regular variation and the spectral measure}\label{subsec:spmeas}

Let $\mathbf{Z}$ be a random vector on $\mathbb{E}:=[0,\infty]^2
\setminus \{(0,0)\}$, with distribution function $F$. The tail of $F$ is \textit{bivariate regularly varying} if there exist a function $b(t)\rightarrow\infty$ and a Radon measure $\nu$ on $\mathbb{E}$, such that
\begin{equation}\label{eq:biregvar}
t\mathbb{P}\left[\frac{\mathbf{Z}}{b(t)}\in\cdot\right]\xrightarrow{v}\nu(\cdot),
\end{equation}
vaguely in $\mathbb{E}$. Notice that this is a straightforward generalization of the univariate case as formulated in Eq. \ref{eq:regvar}.

In terms of dependence structure of the components of $\mathbf{Z}$, it
is often illuminating to consider the equivalent formulation of
Eq. \ref{eq:biregvar} that arises by transforming to polar coordinates.
We define the \textit{polar coordinate} transform of
$\mathbf{Z}=(X,Y)\in\mathbb{E}$ by 
\begin{equation}\label{eq:polar}
(N,\Theta)=POLAR(\mathbf{Z}):=\left(||\mathbf{Z}||,\frac{\mathbf{Z}}{||\mathbf{Z}||}\right),
\end{equation}
where from this point on we use the $L_1$ norm given by $||\mathbf{Z}||=X+Y$.

Bivariate regular variation as formulated in Eq. \ref{eq:biregvar} is
equivalent to the existence of a function $b(t)\rightarrow\infty$ and
a probability measure $\mathbb{S}$ on $\aleph_+$, 
where $\aleph_+=\{\mathbf{z}\in\mathbb{E};||\mathbf{z}||=1\}$,
such that
\begin{equation}\label{eq:polregvar}
\mu_t(\cdot):=t\mathbb{P}\left[\left(\frac{N}{b(t)},\Theta\right)\in\cdot\right]\xrightarrow{v}c\nu_{\gamma}\times\mathbb{S}(\cdot),
\end{equation}
vaguely in $M_+((0,\infty]\times\aleph_+)$. Here
  $\nu_\gamma(r,\infty]=r^{-1/\gamma},\, r>0,$ and $c>0$ and, as
  usual, $M_+((0,\infty]\times\aleph_+)$ are the positive Radon
  measures on $(0,\infty]\times\aleph_+$.
Since there is a
  natural bijection between $\aleph_+$ and $[0,1]$, namely
  $\frac{\mathbf{Z}}{||\mathbf{Z}||}\leftrightarrow\frac{X}{||\mathbf{Z}||}$,
  we can and will assume $\mathbb{S}$ is defined on $[0,1]$. 

The probability measure $\mathbb{S}$ is known as the {\it limit\/} or
\textit{spectral measure}, and it quantifies
the dependence structure among the components of the  bivariate random
vector.  Consider the following two cases which
represent opposite ends of the dependence spectrum \citep{coles:2001,resnickbook:2007}. Suppose $\mathbf{Z}=(X,Y)$ is a random
vector in $\mathbb{E}$ that is bivariate regularly varying as in
Eq. \ref{eq:polregvar}.
\begin{itemize}
\item On one end of the dependence spectrum, if $\mathbf{Z}=(X,Y) $ 
and $X$ and $Y$ are iid,
then $\mathbb{S}$ concentrates on $\theta\in\{0,1\}$,
  corresponding to the axis $x=0$ and $y=0$, respectively. Conversely, if $\mathbb{S}$ concentrates on $\theta\in\{0,1\}$, then there is negligible probability that both $X$ and $Y$ are simultaneously large, and this behavior is called \textit{asymptotic independence}.
\item On the other end of the dependence spectrum, if $\mathbf{Z}=(X,Y)$ and $X=Y$, the two
  components are fully dependent and  $\mathbb{S}$
  concentrates on $\theta=1/2$. This behavior is called
  \textit{asymptotic full dependence}. 
If $\mathbb{S}$
  concentrates on the interior of the $\mathbb{E}$, then we can expect
  $X$ and $Y$ to be highly dependent. 
\end{itemize}

Since $\mathbb{S}$ could be any probability measure, there are infinitely many kinds of dependence structures between the two extreme cases discussed above. Therefore, we focus on the estimation of the spectral measure $\mathbb{S}$ as means of discerning the asymptotic dependence between two random variables with heavy tails.

\subsection{Estimation of the spectral measure $\mathbb{S}$ by the antiranks method.}\label{subsec:antiranks}

For estimating $\mathbb{S}$,  we use the following result. If
$\{\mathbf{Z}_i,i=1,\ldots,n\}$ is a random sample of iid vectors in
$\mathbb{E}$ whose common distribution $F$ is bivariate regularly
varying as in Eq. \ref{eq:biregvar}, then for
$(N_i,\Theta_i):=\left(||\mathbf{Z}_i||,\frac{\mathbf{Z}_i}{||\mathbf{Z}_i||}\right)$: 
\begin{equation}\label{eq:poisstrans}
\frac 1k \sum_{i=1}^n \epsilon_{(N_i/b(\frac nk),\Theta_i)}\Rightarrow c\nu_\gamma\times\mathbb{S},
\end{equation}
as $k\rightarrow\infty,k/n\rightarrow0,$ and $n\rightarrow\infty$. Equation \ref{eq:poisstrans} provides a consistent estimator of $\mathbb{S}$, since
\begin{equation}\label{eq:sdang}
\frac{\sum_{i=1}^n \epsilon_{(N_i/b(\frac nk),\Theta_i)}((1,\infty]\times\cdot)}{\sum_{i=1}^n \epsilon_{N_i/b(\frac nk)}((1,\infty])}\Rightarrow\mathbb{S}(\cdot),
\end{equation}
provided $b(t)=t$. See \cite{huang:1992,
  dehaan:ferreira:2006, resnickbook:2007}.

However, the phrasing of bivariate regular variation in
Eq. \ref{eq:biregvar} requires scaling the two components of $\mathbf{Z}
=(X,Y)$ by the same factor, which implies that
\begin{equation*}
n\mathbb{P}\left[\frac{X}{b_n}\in\cdot\right]\xrightarrow{v}c_1\nu_{\gamma}(\cdot),\quad n\mathbb{P}\left[\frac{Y}{b_n}\in\cdot\right]\xrightarrow{v}c_2\nu_{\gamma}(\cdot),\quad n\rightarrow\infty
\end{equation*}
for $c_j\geq0$ and $j=1,2$. When $c_1>0$ and $c_2>0$, both $X$ and $Y$
have the same shape parameters and their distributions are tail
equivalent \citep{resnick:1971}; this is the \textit{standard regular variation} case.

In practice, we rarely encounter  bivariate heavy tailed data for
which the 
$\gamma$ of each component is the same. For example, consider the
bivariate random vector $(S,D)$. For many decile groups, observe in
Table \ref{tab:hill} that $\gamma_S\not=\gamma_D$. Hence, in order to
estimate $\mathbb{S}$, one possible approach is to
transform the data to the standard case using the antiranks method
\citep{huang:1992,
  dehaan:ferreira:2006, resnickbook:2007}.
This procedure does not
require estimation of the $\gamma$s, yet it achieves transformation to
the standard case, thus allowing  the estimation of
$\mathbb{S}$. However, the transformation destroys the iid property of
the sample and a more sophisticated asymptotic analysis is required. 

We proceed as follows. For iid bivariate data $\{(X_i,Y_i), 1 \leq i \leq
n\}$ from a distribution in a domain of attraction,
define the marginal antiranks by
\begin{equation*}
r_i^{(1)}=\sum_{l=1}^n 1_{\left[X_l\geq X_i\right]},\quad r_i^{(2)}=\sum_{l=1}^n 1_{\left[Y_l\geq Y_i\right]},
\end{equation*}
that is, $r_i^{(j)}$ is the number of $j$th components that are as large as the $j$th component of the $i$th observation.
Then:
\begin{itemize}
\item Transform the data $\{(X_i,Y_i), 1 \leq i \leq n\}$ 
using the antirank transform:
\begin{equation*}
\{\mathbf{Z}_i;1\leq i\leq n\}=\{(k/r_i^{(1)},k/r_i^{(2)});1\leq i\leq n\}.
\end{equation*}
\item  Apply the polar coordinate transformation 
\begin{equation*}
POLAR\left(\frac{k}{r_i^{(1)}},\frac{k}{r_i^{(2)}}\right)=(N_{i,k},\Theta_{i,k}).
\end{equation*}
\item Estimate $\mathbb{S}$ with
\begin{equation}\label{eq:ang}
\hat{\mathbb{S}}_{k,n}(\cdot)=\frac{\sum_{i=1}^n \epsilon_{(N_{i,k},\Theta_{i,k})}((1,\infty]\times\cdot)}{\sum_{i=1}^n \epsilon_{N_{i,k}}((1,\infty])}\Rightarrow\mathbb{S}(\cdot).
\end{equation}
See \cite{resnickbook:2007, dehaan:resnick:1993}.
\end{itemize}

The interpretation of Eq. \ref{eq:ang} is that the empirical probability
measure of those $\Theta$s whose radius $N$ is greater than 1 consistently
approximates $\mathbb{S}$. Hence, we should get a good estimate of
$\mathbb{S}$ by fitting an adequate distribution to the points
$\{\Theta_{i,k};N_{i,k}>1\}$, for a suitable $k$ (see Section \ref{sub:parestspdenssd}). Even though we do not know that
$\mathbb{S}$ has a density, often a density estimate is more striking
than a distribution function estimate. For example, a mode in the
density at $1/2$ reveals a tendency towards asymptotic dependence,
 but modes in the density at 0 and 1 exhibit a tendency towards
 asymptotic independence.

\subsection{Parametric estimation of the spectral density of $(S,D)$}\label{sub:parestspdenssd}

\begin{figure}
\includegraphics{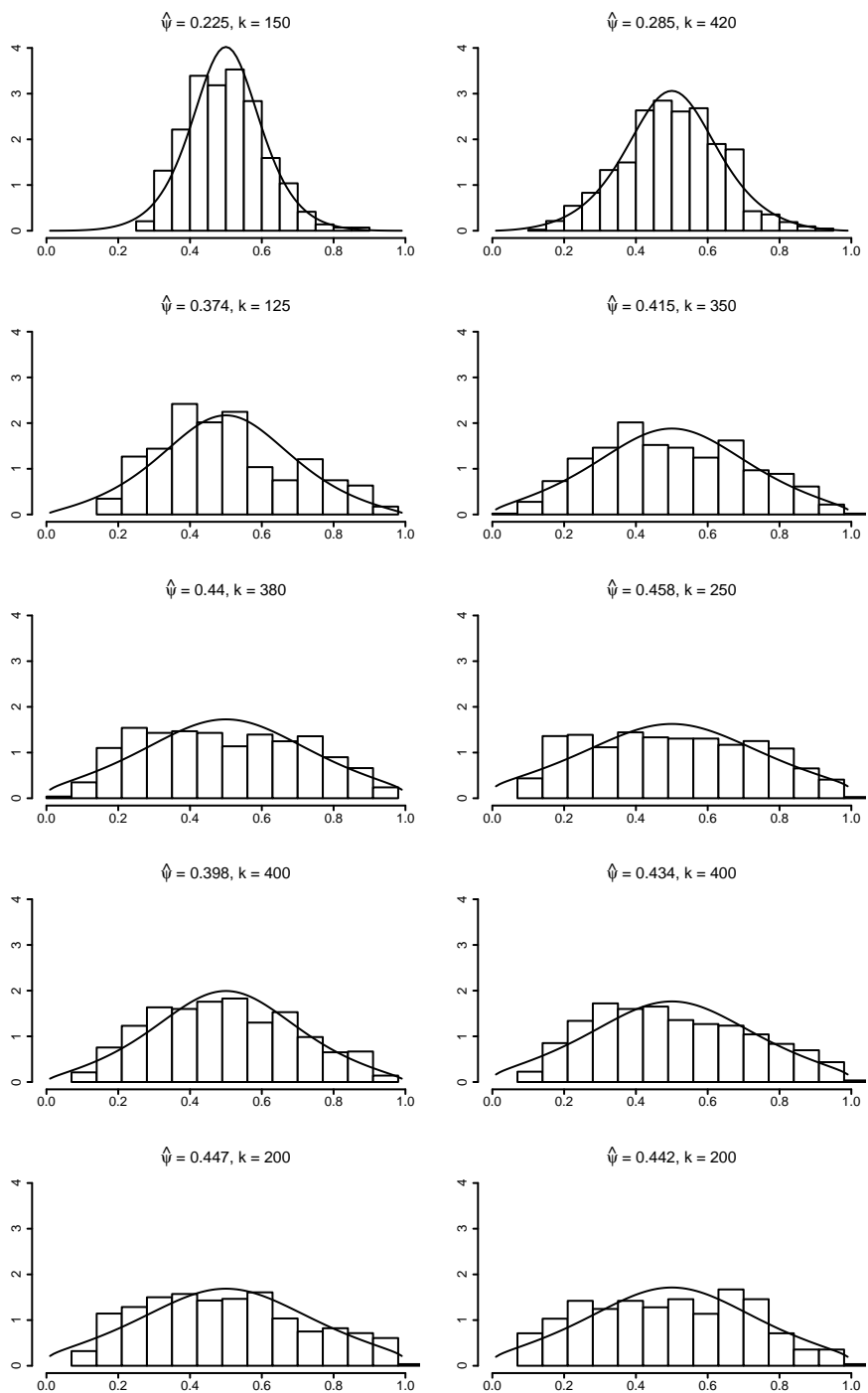}
\caption{Logistic estimates of the spectral density of $(S,D)$ superimposed on the histograms of the points $\{\Theta_{i,k};N_{i,k}>1\}$, starting with the 1st decile group from the upper left and going left to right by row}\label{fig:spdensev}
\end{figure}

Using the antiranks method described above, we transform the points
$\{(S_i,D_i)\}$ for each decile group separately. Figure
\ref{fig:spdensev} shows histograms of the transformed points
$\{\Theta_{i,k};N_{i,k}>1\}$. The histograms suggest that the strength
of the dependence between $S$ and $D$ decreases as $R^\vee$ increases,
since there is increasing mass towards the ends of the interval
$[0,1]$ as the $R^\vee$ goes up. It is certainly apparent that
asymptotic independence does not hold in any decile group.

In order to assess the significance of this apparent trend, we review
a parametric estimator of the spectral density. The histograms in
Fig. \ref{fig:spdensev} show that the spectral density is reasonably
symmetric for each decile group, suggesting that the logistic family
may be an appropriate parametric model. The logistic family is a symmetric model
for the spectral density \citep{coles:2001}, defined by 
\begin{equation}\label{eq:logistic}
h(t)=\frac 12 \left(\frac 1\psi-1\right)t^{-1-\frac{1}{\psi}}(1-t)^{-1-\frac{1}{\psi}}[t^{-\frac{1}{\psi}}+(1-t)^{-\frac{1}{\psi}}]^{\psi-2},\quad 0\leq t\leq1,
\end{equation}
with a single parameter $\psi\in(0,1)$. For $\psi<0.5$, $h$ is unimodal, whereas for increasingly large values of $\psi>0.5$, the density places greater mass towards the ends of the interval $[0,1]$. In fact, asymptotic independence is
obtained as $\psi\rightarrow1$, and perfect dependence is obtained as
$\psi\rightarrow0$. This allows us to quantify the effect of $R^\vee$
on the dependence between $S$ and $D$. 

We first fit the model in Eq. \ref{eq:logistic} to the data $\{\Theta_{i,k};N_{i,k}>1\}$ within each $R^\vee$ decile group by maximum likelihood estimation. The log-likelihood function of $\psi$ based on $t_1,\ldots,t_n$ is
\begin{align}\label{eq:loglike}
l(\psi)=&\sum_{i=1}^n\log\left(\frac{1}{\psi}-1\right)-\sum_{i=1}^n\left(1+\frac{1}{\psi}\right)\log(t_i(1-t_i))\nonumber\\
&+\sum_{i=1}^n(\psi-2)\log(t_i^{-1/\psi}+(1-t_i)^{-1/\psi}),
\end{align}
which we maximize numerically for $0\leq\psi\leq1$. By considering
$\psi$ as a function of $k$, we choose a value of $k$ around which the
estimate of $\psi$ looks stable. Figure \ref{fig:spdensev} shows that
the logistic estimates of the spectral density are in close agreement
to the histogram of the points. On top of each plot, we indicate the
maximum likelihood estimates of $\psi$ and the choice of $k$ in the
corresponding decile group. The estimates of $\psi$ confirm a decline
in dependence between $S$ and $D$ as the decile group increases, as
measured by increasing estimates of $\psi$.

\begin{figure}[tb]
\includegraphics{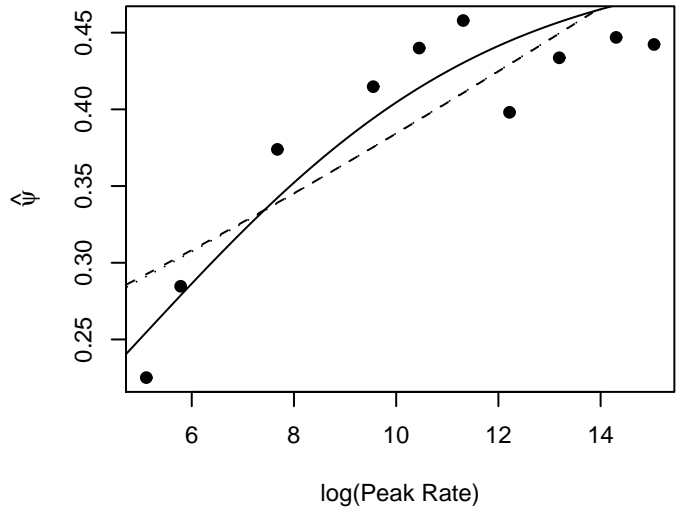}
\caption{Parameter $\psi$ as a function of $\log(R^\vee)$ and three linear models of the form given by Eq. \ref{eq:linearpsi} superimposed: (solid line) link function in Eq. \ref{eq:halflogit}, (dashed line) logit link, (dotted line) probit link. The logit and probit links are almost indistinguishable in the range of the data}\label{fig:linearpsi}
\end{figure}

We now study the form of this decline by fitting a global trend model
simultaneously to all the peak rate decile groups, using the same data
{(antirank transformed, polar coordinate
  transformed, thresholded)}
employed for the separate analyses. In this joint study, the parameter $\psi$ in Eq. \ref{eq:loglike}
is a function of $R^\vee$ as follows:
\begin{equation}\label{eq:linearpsi}
g^{-1}(\psi)=\beta_0 + \beta_1 \log(R^\vee),
\end{equation}
where $g$ is a link function. The used of $\log(R^\vee)$ instead of $R^\vee$ is a common technique in linear models to improve fit. Since $\psi\in(0,1)$, natural choices of $g$ are the logit and the probit functions. However, as shown in Fig. \ref{fig:linearpsi}, the link function
\begin{equation}\label{eq:halflogit}
g(x)=\frac{0.5}{1+e^{-x}}
\end{equation}
is more adequate than the usual logit
  or probit links. The link given by Eq. \ref{eq:halflogit} is very
similar to the logit link, but confines the possible values of $\psi$
to the interval $(0,0.5)$ {and is suggested by the
  fact that in}
Fig. \ref{fig:spdensev} the histograms of the points
$\{\Theta_{i,k};N_{i,k}>1\}$ put all mass around an apparent mode at
$0.5$. This behavior corresponds to $\psi<0.5$ as previously
stated. 

Figure \ref{fig:linearpsi} exhibits in various ways the logistic
parameter $\psi$ as a function of peak rate {using Eq. \ref{eq:linearpsi}.} First, we plot the points
$\mathcal{P}=\{(med^{(i)}, \hat{\psi}^{(i)});1\leq i\leq10\}$, where $med^{(i)}$ is the
median of the  $\log R^\vee$ {variable for sessions in the $i$th decile
group} and $\hat{\psi}^{(i)}$ is the {maximum likelihood estimated} logistic
parameter in the $i$th decile group. In Fig. \ref{fig:linearpsi}, we 
superimpose on $\mathcal{P}$ the estimated Eq. \ref{eq:linearpsi} using the link function in
Eq. \ref{eq:halflogit}, showing that the goodness of fit of the
model in Eq. \ref{eq:linearpsi} {is quite reasonable}. 

To assess the effect of $R^\vee$ on the dependence structure of
$(S,D)$, we  focus on $\hat{\beta_1}$. Observe that
Eq. \ref{eq:loglike} gives the log-likelihood of the model for independent
observations. Since $\{\Theta_{i,k};N_{i,k}>1\}$ is not an independent
sample due to the antirank transform, the classical maximum likelihood
theory is not strictly applicable.
Hence, to  quickly compute the standard
error of $\hat{\beta_1}$ we bootstrap the whole model. However,
several authors have shown in the context of heavy-tailed phenomena
that if the original sample is of size $n$, then the bootstrap sample
size $m$ should be of smaller order for  asymptotics to work as
desired \citep{athreya:1987b, deheuvels:mason:shorack:1993,
gine:zinn:1989,
hall:1990,
resnickbook:2007}. In connection with the estimation of the spectral
measure, the bootstrap procedure works as long as $m\to\infty$,
$m/n\to0$ and $n\to\infty$. 
Therefore, a bootstrap procedure to estimate
the standard error of $\hat{\beta}_1$ is constructed as follows: 
\begin{enumerate}[(i)]
\item From the original sample $\{(S_i,D_i,R_i^\vee);1\leq i\leq 44136\}$, a bootstrap sample $\{(S_i^*,D_i^*,R_i^{\vee*});1\leq i\leq 10000\}$ is obtained. Notice that the bootstrap sample size is of smaller order than the original sample size. Our choice of $m=10000$ owes to the need of having enough data points to perform estimation. However, the choice of the bootstrap sample size is as tricky as choosing the threshold $k$ used in, say, Hill estimation. Hence, this may be subject of further study.
\item Split the bootstrap sample $\{(S_i^*,D_i^*,R_i^{\vee*});1\leq i\leq 10000\}$ into 10 groups according to the quantiles of $R_i^{\vee*}$.
\item Within each bootstrap decile group, transform the data $\{(S_i^*,D_i^*);1\leq i\leq1000\}$ using the antirank transform and then transform to polar coordinates to obtain $\{\Theta_{i,k}^*;N_{i,k}^*>1\}$. Here, for each bootstrap decile group we use the same value of $k$ that is used in the original estimation. These values are shown in Fig. \ref{fig:spdensev}.
\item Fit the global linear trend simultaneously to all the bootstrap decile groups, by maximizing Eq. \ref{eq:loglike} with $\psi$ as a function of $R^{\vee*}$ as in Eqs. \ref{eq:linearpsi} and \ref{eq:halflogit}. Hence, we obtain a bootstrap replication $\hat{\beta}_{1,b}^*$.
\item Repeat steps (i)-(iv) $B=1000$ times and estimate the standard error of $\hat{\beta_1}$ by the sample standard deviation of the $B$ replications
\begin{equation}\label{eq:btrapse}
\widehat{se}(\hat{\beta}_1)=\left\{\frac{1}{B-1}\sum_{b=1}^B[\hat{\beta}_{1,b}^*-\hat{\beta}_1^*]^2\right\}^{1/2},
\end{equation}
where $\hat{\beta}_1^*=\sum_{b=1}^B\hat{\beta}_{1,b}^*/B$.
\end{enumerate}

Table \ref{tab:linearpsi} summarizes the estimated parameters of the
linear model for $\psi$ and their standard errors. Recall that our model assesses
dependence through the value of $\psi$. From these results,
we conclude that there is a significant effect of the level of
$R^\vee$ in the dependence structure of $(S,D)$, since $\hat{\beta_1}$
is significantly different from 0. 

We conclude that
Eq. \ref{eq:linearpsi} jointly with Eq. \ref{eq:halflogit} provide an
adequate description of the behavior of $\psi$ across the decile
groups. 

\begin{table}
\caption{Summary of estimated linear model given by Eqs. \ref{eq:linearpsi} and \ref{eq:halflogit}}
\begin{tabular}{l c r c r}
\hline\noalign{\smallskip}
 & & Estimated parameter & & Bootstrap standard errors\\\hline\noalign{\smallskip}
$\hat{\beta_0}$ & & -1.432 & & 0.219 \\
$\hat{\beta_1}$ & & 0.288 & & 0.127\\\hline\noalign{\smallskip}
\end{tabular}
\label{tab:linearpsi}
\end{table}

\subsection{Parametric estimation of the spectral density of $(R,S)$ and $(R,D)$}\label{subsec:remind}

\begin{figure}[b]
\includegraphics{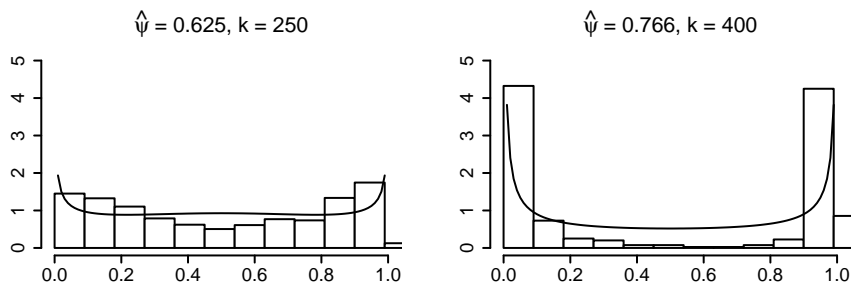}
\caption{Logistic estimates in the 10th decile group superimposed on the histograms of the points $\{\Theta_{i,k};N_{i,k}>1\}$ \textit{Left} Spectral density of $(R,S)$ \textit{Right} Spectral density of $(R,D)$ }\label{fig:spdenscev}
\end{figure}

We now transform the points $\{(R_i,S_i)\}$ and the points $\{(R_i,D_i)\}$ in the 10th decile group using the previously described antirank transform. Figures \ref{fig:spdenscev}(a) and \ref{fig:spdenscev}(b) exhibit histograms of the transformed points $\{\Theta_{i,k};N_{i,k}>1\}$ corresponding to the pairs $(R,S)$ and $(R,D)$, respectively. Both histograms look reasonably symmetric, and thus the modeling is done via the logistic family defined by Eq. \ref{eq:logistic}.

Figure \ref{fig:spdenscev} shows that the fitted logistic models are in close agreement with the empirical distribution of the points $\{\Theta_{i,k};N_{i,k}>1\}$. Notice that the parameter $\psi$ of the logistic density corresponding to the pair $(R,D)$ is closer to 1 than the parameter $\psi$ of the density corresponding to the pair $(R,S)$. This suggests that for the group of sessions with the highest values of $R^\vee$, the scheme $RD$ (in which $R$ and $D$ are independent, at least asymptotically), is more adequate than the scheme $RS$ (in which $R$ and $S$ are  independent, at least asymptotically). This conclusion is exactly the opposite to \cite{sarvotham:riedi:baraniuk:2005}'s, since they recommend using the scheme $RS$ for the group with the highest peak rates (that is, their alpha group).

The fact that for the sessions with the highest values of peak rate
$R^\vee$,
{we have  $(R,D)$ close to asymptotically independent may have the
following interpretation.}
 Users with high bandwidth pay little or no
attention to the duration of their downloads; this is expected because
such users know that probably their lines are capable of downloading
any file, no matter how long it takes. 

\section{Dependence structure of $(S,D,R)$ when $R$ does not have heavy tails}\label{sec:depcev}

We now investigate the dependence structure of $(R,S)$ and $(R,D)$ in
the first nine decile groups, that is, those with values of $R^\vee$
in the decile ranges $(10(g-1)\%,10g\%],$ $g=1,\ldots,9$. For these
groups, there is evidence that the distribution of $R$ is not heavy
tailed. Moreover, the diagnostics in Section \ref{sub:excess} suggest
that $R\not\in\mathcal{D}(G_\gamma)$ for any
$\gamma\in\mathbb{R}$. However, the other variables $S$ and $D$ have
heavy tails in these decile groups, and we can make use of the
conditional extreme value model \citep{
heffernan:tawn:2004,
heffernan:resnick:2007,
das:resnick:2008b,
das:resnick:2008c}
to study the dependence structure of
the pairs $(R,S)$ and $(R,D)$.

\subsection{The conditional extreme value model}\label{sub:cev}

Classical bivariate extreme value theory assumes that both variables
are in some maximal domain of attraction. When one variable is in a
domain of attraction, but the other is not, the conditional extreme
value model, or CEV  provides a candidate model. 

Let $\mathbf{Z}=(X,Y)\in\mathbb{E}=[0.\infty]^2\setminus\{(0,0)\}$ and let $\bar{\mathbb{E}}^{(\gamma)}$ be the right closure of $\mathbb{E}^{(\gamma)}=\{y\in\mathbb{R}:1+\gamma y>0\}$. The CEV model assumes that $Y\in\mathcal{D}(G_\gamma),\gamma\in\mathbb{R},$ with normalizing sequences $a(t)>0$ and $b(t)$ as in Eq. \ref{eq:mdab}. In addition, the CEV model assumes that there exist functions $\alpha(t)>0$, $\beta(t)\in\mathbb{R}$ and a non-null Radon measure $\mu$ on the Borel subsets of $[-\infty,\infty]\times\bar{\mathbb{E}}^{(\gamma)}$ such that the following conditions hold for any $y\in\mathbb{E}^{(\gamma)}$:
\begin{enumerate}[(i)]
\item For $\mu-$continuity points $(x,y)$: 
\begin{equation*}
t\mathbb{P}\left(\frac{{X}-\beta(t)}{\alpha(t)}\leq x,\frac{Y-b(t)}{a(t)}>y\right)\rightarrow\mu([-\infty,x]\times(y,\infty]),\quad t\rightarrow\infty.
\end{equation*}
\item $\mu([-\infty,x]\times(y,\infty])$ is not a degenerate distribution in $x$.
\item $\mu([-\infty,x]\times(y,\infty])<\infty$.
\item $H(x):=\mu([-\infty,x]\times(0,\infty])$ is a probability distribution.
\end{enumerate}

The reason for the name CEV is that, assuming $(x,0)$ is a $\mu-$continuity point:
\begin{equation}\label{eq:cev}
\mathbb{P}\left(\left.\frac{X-\beta(t)}{\alpha(t)}\leq x\right|Y>b(t)\right)\rightarrow H(x),\quad t\rightarrow\infty.
\end{equation}
Therefore, Eq. \ref{eq:cev} provides a way to study the dependence structure of the components of $\mathbf{Z}$ when only one is in a maximal domain of attraction.

\subsection{Checking the CEV model}\label{seb:checkcev}

We now review a method for checking the adequateness of the CEV model, recently developed by \cite{das:resnick:2008b}. Suppose $\{(X_i,Y_i);1\leq i\leq n\}$ are\textcolor{blue}{???} iid from the CEV model. Define:
\begin{itemize}
\item $Y_{(1)}\geq\ldots,Y_{(n)}$: The upper-order statistics of $Y_1,\ldots,Y_n$.
\item $X_i^*,1\leq i\leq n$: The $X$-variable corresponding to $Y_{(i)}$, also called the \textit{concomitant} of $Y_{(i)}$.
\item $r_{i,k}^*=\sum_{l=1}^k 1_{[X_l^*\leq X_i^*]}$: The rank of $X_i^*$ among $X_1^*,\ldots,X_k^*$.
\end{itemize}

The \textit{Hillish statistic} of $\{(X_i,Y_i);1\leq i\leq n\}$ is defined as
\begin{equation*}
\mathrm{Hillish}_{k,n}:=\frac 1k \sum_{j=1}^k \log \frac{k}{r_{i,k}^*}\log \frac kj.
\end{equation*}
Under $H_0 : \{(X_i,Y_i);1\leq i\leq n\}$ \textit{are iid from a CEV
  model}, \citet{das:resnick:2008c}  proved that as
$k\rightarrow\infty,k/n\rightarrow0,$ and $n\rightarrow\infty$: 
\begin{equation}\label{eq:hillish}
\mathrm{Hillish}_{k,n}\xrightarrow{P} I_{\mu,H},
\end{equation}
where $I_{\mu,H}$ is a constant that depends on $\mu$ and $H$ defined in Section \ref{sub:cev}.

Like the Hill estimator, the Hillish statistic depends on the number $k$, so we make a \textit{Hillish plot} $\{(k,\mathrm{Hillish}_{k,n});k\geq1\}$ and observe whether the plot has a stable regime. If that is the case, we conclude that the CEV model is adequate for $(X,Y)$.

\subsection{Checking the CEV model for $(R,S)$}\label{sub:checkcevrs}

\begin{figure}[htbp]
\includegraphics{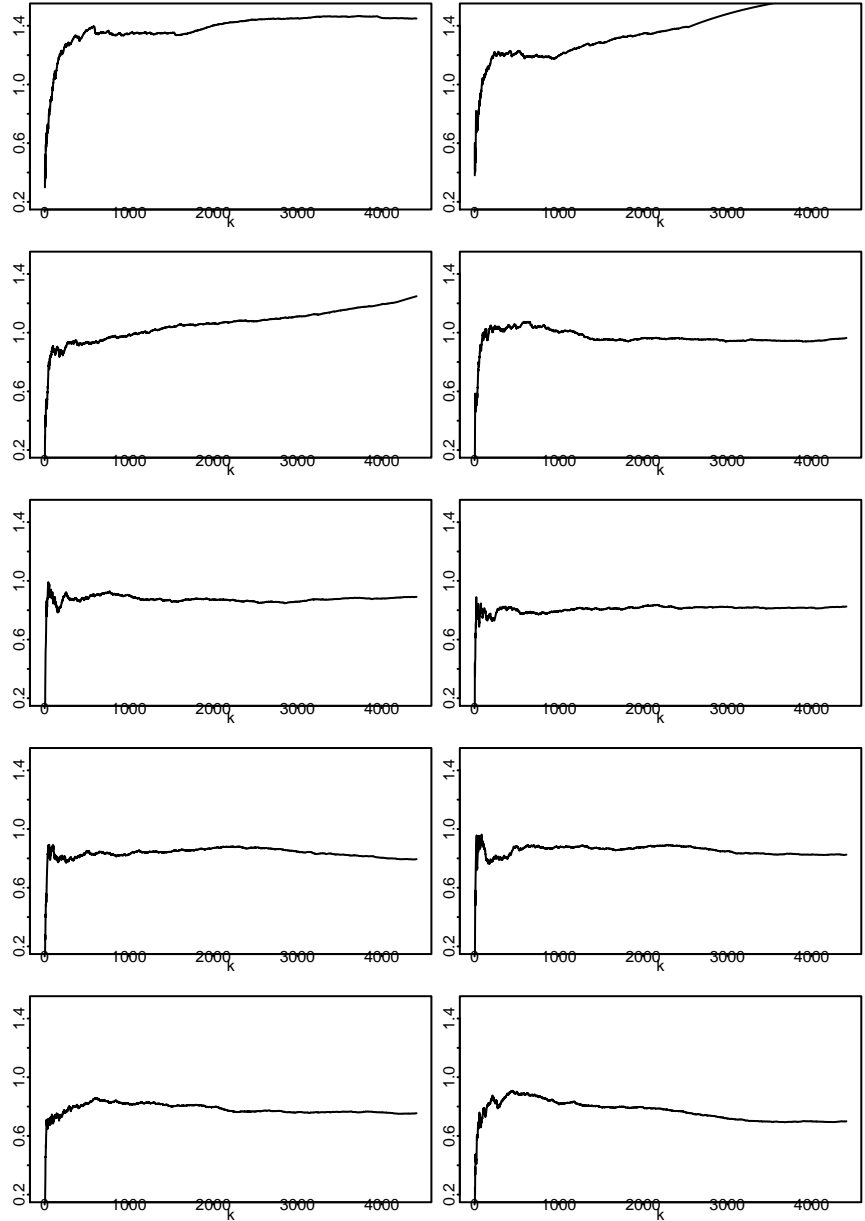}
\caption{Hillish statistic of $(R,S)$, starting with the 1st decile group from the upper left and going by row}\label{fig:hillish}
\end{figure}

The CEV model appears as a candidate model for $(R,S)$ or $(R,D)$ for
any  one of the lowest 9 decile groups, since for these groups
$R$ does not appear to be in a domain of attraction, while both $S$
and $D$  have heavy tails. We found that the CEV model is
adequate for $(R,S)$ within each of the lowest nine $R^\vee$-decile
groups. Here we present our results. 

Figure \ref{fig:hillish} shows Hillish plots for checking the CEV
model for $(R,S)$ in the lowest nine decile groups. Apart from the
second and third decile groups, all the plots look exceptionally
stable. Although the plots do not look as good for the second and
third decile groups, we still find a stable regime about the $k=800$
upper order statistic, which supports the CEV model for these groups
as well. This emphasizes that more detailed structure exists for the beta group of \cite{sarvotham:riedi:baraniuk:2005} and that further segmentation reveals more information.

Moreover, observe that the limit constant $I_{\mu,H}$ varies with the decile group. In effect, $I_{\mu,H}$ decreases as $R^\vee$ goes up. Since $I_{\mu,H}$ depends on the limit $H$ in Eq. \ref{eq:cev}, this suggests that the conditional distribution of $R$ given $S$ varies with the decile group. Hence, the dependence structure of $(S,D,R)$ depends on the explicit level of $R^\vee$.

In addition, the Hillish plots reject the CEV model for the pair (R,D)
in all the decile groups. {We have not displayed these plots.}

\section{The Poisson property}\label{sec:poiss}

There is considerable evidence against the Poisson model as the
generating mechanism for network traffic at the packet-level
\citep{
paxson:floyd:1995,
willinger:taqqu:sherman:wilson:1997,
willinger:paxson:1998,hohn:veitch:abry:2003}. However, the classical mechanism of
Poisson arrival times, namely human activity generating \textit{many
  independent} user connections to a server each with \textit{small
  probability} of occurrence, is still present in the network traffic
at higher levels. A significant example is provided by
\cite{park:shen:marron:hernandez-campos:veitch:2006}, which shows that
``navigation bursts" in the server occur according to the Poisson
model. 

Here we found that, although the Poisson model does not appear to activate the overall network traffic, it does initiate user sessions  for any given group of sessions whose peak rate $R^\vee$ is in a fixed inter-decile range. This allows for quite straightforward simulation within each decile group via a homogeneous Poisson process.

Recall we split the sessions into 10 groups according
to the deciles of $R^\vee$. For any given decile group, suppose that
$\Gamma_i$ are the starting times of the user sessions in increasing
order; if necessary, we relabel sessions within the group.
Let $\Delta_i=\Gamma_{i+1}-\Gamma_i$ be the session interarrival
times. A homogeneous Poisson process is characterized by
$\{\Delta_i\}$ being iid with the exponential $\exp(\lambda)$ as the
common distribution function, for some parameter $\lambda>0$. 

\subsection{Checking the exponential distribution for interarrival times}\label{sub:poissexp}

\begin{figure}[htb]
\centering
\includegraphics{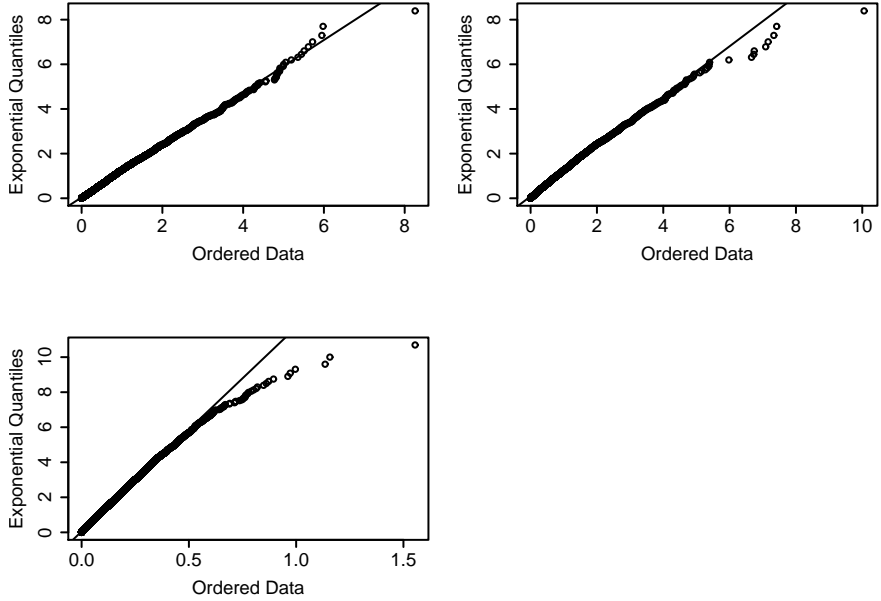}
\caption{Exponential QQ plots of the interarrival times of sessions \textit{Upper left} 4th decile group \textit{Upper right} 10th decile group \textit{Lower left} overall traffic}\label{fig:qqexp}
\end{figure}

We first check that $\{\Delta_i\}$ may be accurately modeled as exponential random variables within each $R^\vee$ decile group.
As examples, Fig. \ref{fig:qqexp} \textit{Upper left} and textit{Upper right} panels exhibit
exponential QQ plots for $\{\Delta_i\}$ for the 4th and 10th decile
groups, respectively, which compare the quantiles of the empirical and
theorical distributions. It is striking how well a straight line trend
is shown, and this result replicates across all the decile
groups. However, when all the sessions are put together in a single 
population, the session interarrival times have right tails noticeably
heavier than exponential, as Fig. \ref{fig:qqexp} \textit{Lower left} shows.

Interestingly, we found that the interarrival times within each decile group are not exponentially distributed when rather than using the deciles of $R^\vee$ for the segmentation, we use the deciles of any of the two previous predictors of burstiness, namely $I_\delta$ and $R_\delta$.

\subsection{Checking the independence of interarrival times}\label{sub:poissind}

Within each decile group, can we use the independence model for
$\{\Delta_i\}$?
We investigated this question with  the sample
autocorrelation function. 
The \textit{sample autocorrelation function (acf)} of $\Delta_1,\ldots,\Delta_n$ at lag $h$ is defined as
\begin{equation*}
\hat{\rho}(h)=\frac{\sum_{i=1}^{n-h}(\Delta_i-\bar{\Delta})(\Delta_{i+h}-\bar{\Delta})}{\sum_{i=1}^n(\Delta_i-\bar{\Delta})^2}.
\end{equation*}

In Section \ref{sub:poissexp}, we showed $\{\Delta_i\}$ can be accurately modeled as an exponential random sample, and thus we can safely assume that the variances of $\Delta_i$ are finite. Therefore, the standard $L_2$ theory applies and Bartlett's formula from classical time series analysis \citep{brockwell:davis:1991} provides asymptotic normality of $\hat{\rho}(h)$ under the null hypothesis of independence, namely,
\begin{equation}\label{eq:normacf}
\sqrt n  \hat{\rho}(h)\xrightarrow{d}N(0,1).
\end{equation}

Based on Eq. \ref{eq:normacf}, we can test $H_0:\{\Delta_i\}$
\textit{are independent} by first determining the corresponding
(upper) quantile $z_{1-\alpha}$ of the normal distribution, and then
plotting the sample acf as a function of the lag $h$. According to
Eq. \ref{eq:normacf}, approximately $1-\alpha$ of the points
$\hat{\rho}(h)$ should lie between the bounds $\pm
z_{1-\alpha}n^{-1/2}$, and, if so, there is no evidence against $H_0$.

\begin{figure}[tb]
\includegraphics{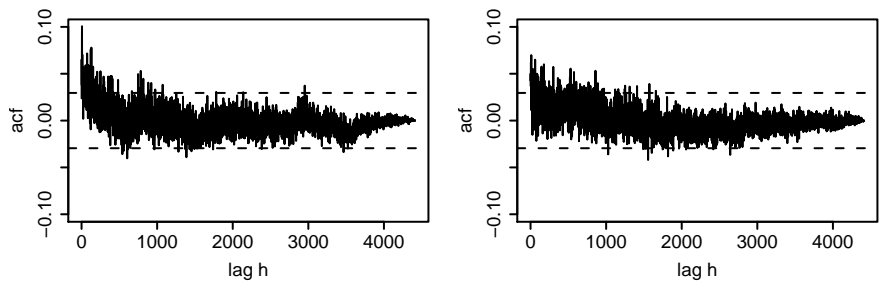}
\caption{Sample autocorrelation functions of $\Delta_i$ \textit{Left} 4th decile group \textit{Right} 10th decile group}\label{fig:acf}
\end{figure}

Figure \ref{fig:acf} \textit{Left} and \textit{Right} panels exhibit sample acf plots
for $\{\Delta_i\}$  for  the 4th and 10th decile groups. In 
each figure, we plot the confidence bounds for an $\alpha=0.05$. We
counted 178 and 141 ``spikes" coming out of those bounds,
respectively. This represent less than $5\%$ of the total of 4414. In
general, we found that less than $5\%$ of the spikes lie outside the
bounds for all the decile groups. Based on the sample acf,
there is no evidence
against the independence of $\{\Delta_i\}$ within each decile.

\section{Final remarks and conclusions}\label{sec:final}

For the purposes of illustration, we have presented our analysis on a
data set publicly available as of May 2009 at \url{http://pma.nlanr.net/Special/index.html} through the National
Laboratory for Applied Network Research (NLANR). The particular data
file chosen for the analysis is ``19991207-125019", which can be found
within a collection of network data traces dubbed \textit{Auckland
  II} recorded in 1999. We have successfully tested our analyses and proposed models given by Eqs. \ref{eq:logistic} and \ref{eq:linearpsi} for other data files in the collection \textit{Auckland II},
as well as a more recent collection dubbed \textit{Auckland VIII} recorded in 2003.

Unfortunately, the NLANR website will be shutdown in May 2009. However, the Waikato Internet Traffic Storage or WITS (\url{http://www.wand.net.nz/wits/}) has expressed a hope to be able to make some of the NLANR's data sets (in particular, those corresponding to the Auckland's series) freely available for researchers to download in the near future.

The reason for our choice of the logistic family and the 
linear trend is because they allow for a simple description of the dependence structure of $(S, D)$ via the logistic parameter.
As depicted in Figs.
\ref{fig:spdensev} and \ref{fig:linearpsi}, the proposed logistic
model defined by Eq. \ref{eq:logistic}, jointly with the linear trend as in
Eq. \ref{eq:linearpsi}, does a sound job of explaining the dependence
structure of $(S,D)$ as a function of the explicit level of the peak
rate $R^\vee$. 

Our findings can yield more accurate simulation methods for network
data. The following is an outline of a procedure to simulate network
data traces: 
\begin{enumerate}
\item Bootstrap from the empirical distribution of $R^\vee$ and split the data into, say, 10 groups according to the empirical deciles.
\item Conditionally on the decile group, simulate the starting times
  $\Gamma$ of the sessions via a homogeneous Poisson process. This
  means that the Poisson rate depends on the decile group; for
  example, from the original data set  estimate the Poisson rates for
  each decile group and use them here.
\item From $R^\vee$, compute $\psi$ using the estimated linear trend as in Eq. \ref{eq:linearpsi} and use it to simulate an ``angle" $\Theta$.
\item Simulate the radial component $N$ as a heavy tailed random
  variable, for instance the Pareto \citep{resnickbook:2007, dehaan:resnick:1993}.
\item Finally, transform $(N,\Theta)$ to Cartesian coordinates in order to get $(S,D)$ and compute $R=S/D$.
\end{enumerate}
We are considering details of a software procedure to implement this simulation suggestion.

Our analyses can be readily extended to other segmentation
schemes. For instance, 
heterogeneous traffic comprising different types of applications undoubtedly
behaves differently from more homogeneous traffic,
a fact used to justify the modeling in 
\cite{dauria:resnick:2008}.
Our
analyses should provide useful insights by investigating the
dependence structure according to the application type.

On another direction, we have shown evidence for the two following models:
\begin{itemize}
\item The classical extreme value theory for the pair $(S,D)$, in which both components are heavy-tailed.
\item The conditional extreme value (CEV) model for the pair $(R,S)$,
  in which only one component, namely $S$, is heavy-tailed. 
\end{itemize}
Given the fact that $R=S/D$, we are investigating the conditions on
the CEV model for $(R,S)$ that imply the classical model for $(S,D)$,
and vice versa. 

\section{Acknowledgement}
Early discussions with Janet Heffernan significantly shaped the
directions of these investigations and our final effort reflects the
benefit of  her initial creative inputs. In particular the idea of
using link functions in Section \ref{subsec:remind} was hers.

\bibliographystyle{plainnat}
\bibliography{bibfile}

\end{document}